\documentclass[aps,prl,twocolumn,groupedaddress,showpacs]{revtex4}

\usepackage{graphicx}
\newcommand{\comment}[1]{}

\begin{document}


\title{Ion specificity and anomalous electrokinetic effects in hydrophobic nanochannels}


\author{David M. Huang, C\'{e}cile Cottin-Bizonne, Christophe Ybert and Lyd\'{e}ric Bocquet}
\email{lyderic.bocquet@univ-lyon1.fr}
\affiliation{
Laboratoire de Physique de la Mati\`ere Condens\'ee et Nanostructures,
Universit\'e Lyon 1, CNRS, UMR 5586, Villeurbanne, F-69622, France}


\date{\today}

\begin{abstract}
We demonstrate with computer simulations that anomalous electrokinetic
effects, such as ion specificity and non-zero zeta potentials for uncharged
surfaces, are generic features of electro-osmotic flow in hydrophobic channels.
This behavior is due to the stronger attraction of larger ions to the
``vapour--liquid-like'' interface induced by a hydrophobic surface.
An analytical model involving a modified Poisson--Boltzmann description for the ion
density distributions is proposed, which allows the anomalous flow profiles to be predicted
quantitatively. This description incorporates as a crucial component
an ion-size-dependent hydrophobic solvation energy. These results provide
an effective framework for predicting specific ion effects, with important implications for 
the modeling of biological problems.
\end{abstract}


\maketitle


Hydrophobic surfaces are at the origin of many surprising and
potentially useful effects \cite{Chandler}, such as hydrodynamic slippage
at hydrophobic surfaces \cite{Lauga,Joly} and the formation of nanobubbles at the surface \cite{Attard}.
A common feature underlying many of these phenomena is the
formation of a layer of depleted water density near the
surface~\cite{netz04} -- a vapor layer in the case of extremely
hydrophobic surfaces. Vapor--liquid interfaces have been found in
recent spectroscopic experiments~\cite{ghosal05} and computer
simulations~\cite{vrbka04} to attract large and polarizable ions such
as bromide and iodide, but not small ions like sodium and
chloride. This ion-specific behavior, contrary to traditional
theories of electrolyte interfaces, which only take into account
differences in ion valency~\cite{hunter01}, is behind the
substantial dependence on anion type of the surface tension of
aqueous solutions of halide salts~\cite{bostrom05}.
Just as ion specificity affects equilibrium properties of
vapor--liquid interfaces like surface tension, a similarly
important role is expected for dynamic phenomena near the
``vapor--liquid-like'' interfaces induced by hydrophobic surfaces.
The implications are considerable for fluid transport in
microfluidic (``lab-on-chip'') devices~\cite{squires05}, for which
surface effects are predominant and electrokinetic techniques for
driving flows widely used, but also for the modeling of biological systems \cite{Chandler}, 
for which ion-specific Hofmeister series are ubiquitous \cite{Ninham}.

In this work, we investigate by
computer simulations the anomalous electrokinetic effects that
arise in electro-osmotic (EO) flow through hydrophobic channels
due to interfacial ion specificity. Furthermore  we develop a simple
model, comprising continuum hydrodynamic equations and a
modified Poisson--Boltzmann (PB) description for the ion density
distributions. Remarkably, our theory is able to predict accurately the effects of ion 
specificity on the simulated EO flow profiles and zeta potential, pointing
furthermore to the crucial role of the hydrophobic solvation
energy. Our analytic theory is a powerful tool for describing
specific ion effects and their consequences on dynamics.

The system studied comprised a solution of monoatomic, monovalent salt ions
in water, confined between two parallel solid walls.
A total of 2160 fluid molecules were used in
all cases and the SPC/E simple point charge model was
employed for the aqueous solvent.
Each wall was composed of 648 atoms arranged
in three unit cell layers of an fcc lattice oriented in the
$\langle 100 \rangle$ direction
(lateral dimensions $(x,y)$: 48.21~\AA\ $\times$ 32.14~\AA).
To model charged surfaces, identical charges were
added to each of the atoms in the top solid layer in contact
with the fluid. 
The inter-wall distance was adjusted such that
the average pressure, defined by the force per unit area on the solid
atoms, was approximately 10~atm in equilibrium simulations.
Periodic boundary conditions were applied in the
$x$ and $y$ directions, while empty space was added in the $z$ direction
such that the total system was three times as large as the primary
simulation cell, which was centered at $z=0$.

Simulations were carried out with the \texttt{LAMMPS}~\cite{lammps}
molecular dynamics package. Bond length and angle constraints for the
rigid water molecules were enforced with the \texttt{SHAKE}
algorithm and a constant temperature of 298~K was maintained
with a Nos{\'e}--Hoover thermostat (applied only to degrees
of freedom in the $y$ direction in the flow simulations where flow is along $x$).
Electrostatic interactions were calculated with the
particle--particle particle--mesh (PPPM) method,
with a correction applied to remove the
dipole--dipole interactions between periodic replicas in the $z$ direction.
Short-ranged van der Waals interactions between particles
were modeled with the Lennard-Jones (LJ) potential,
$v_{ij} (r_{ij}) = 4\varepsilon_{ij}\left[\left(\sigma_{ij}/r_{ij}\right)^{12}
- \left(\sigma_{ij}/r_{ij}\right)^{6} \right]$ for an interparticle
separation of $r_{ij}$ and particle types $i$ and $j$
($\sigma_{ij} = \left( \sigma_{ii} + \sigma_{jj}\right)/2$
and $\varepsilon_{ij} = \sqrt{\varepsilon_{ii}\varepsilon_{jj}}$).
All LJ interactions were truncated and shifted to zero at 10~\AA.
For the solid atoms, LJ parameters were chosen
to create a physically reasonable, albeit idealized, surface:
we took $\sigma_{\text{ss}} = 3.37$~\AA, used a close-packed
density, $\rho_{\text{s}} = \sigma_{\text{ss}}^{-3}$, and
chose $\varepsilon_{\text{ss}} =$ 0.164 and 2.08~kcal/mol
to create, respectively, a hydrophobic
and a hydrophilic surface, as characterised by the
contact angles of a water droplet on these surfaces
of roughly 140 and 55$^\circ$.

\begin{figure}
\includegraphics*[width=8.0cm]{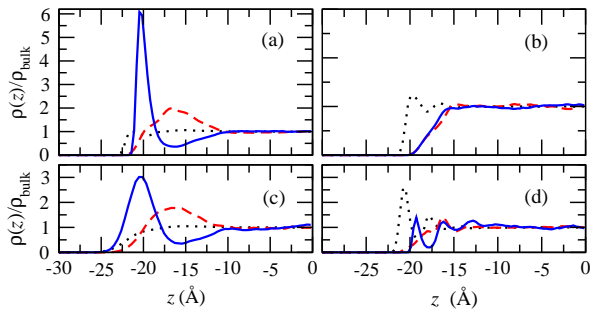}
\caption{\label{fig:ion_density}Simulated density profiles of
negative ions (solid lines), positive ions (dashed lines), and water (dotted lines) 
 for roughly 1-M solutions of: (a) NaI and (b) NaCl between
neutral hydrophobic surfaces; (c) NaI at liquid--vapor interface; (d) 
NaI between neutral hydrophilic surfaces.}
\vspace{-0.5cm}
\end{figure}

EO flow of solutions of either NaI or NaCl were studied, the only
difference between the two cases being anion size. Except for one
case, we used ion LJ parameters from Ref.~\cite{koneshan98}: we
chose $\sigma_{ii} = 6.00$~\AA\ for I$^-$ (instead of 5.17~\AA) to
reproduce approximately liquid--vapor interfacial ion densities
measured in simulations of NaI/water solutions of similar
concentration but using more complex polarizable force
fields~\cite{vrbka04}. Our simulated density profiles are shown in
Fig.~\ref{fig:ion_density}c. Although simulations have shown that
ion polarizability plays a role in stabilizing I$^-$ at the
air--water interface~\cite{vrbka04}, the dominant contribution to
the stabilization free energy is associated with the solvation
energy~\cite{archontis06}, which is accounted for in our simulations.
Thus, we regard our simple
parametrization of the iodide LJ diameter, coupled with the use of
non-polarizable force fields, as adequate for the purpose of capturing
the dynamic consequences of the experimentally observed surface
enhancement of I$^-$.

The effects of anion size and surface wettability on interfacial ion
densities are illustrated in Fig.~\ref{fig:ion_density}.
While Cl$^-$ is not found near the hydrophobic surface
in Fig.~\ref{fig:ion_density}b, Fig.~\ref{fig:ion_density}a shows a
substantially enhanced interfacial I$^-$ concentration.
No such enhancement is seen for I$^-$ ions near
the hydrophilic surface (Fig.~\ref{fig:ion_density}d) even though
the direct ion--solid interactions are stronger in this case, indicating
that the ion density profiles arise largely due to the
water structure induced by the surface.

EO flow was induced in our simulations by applying an
electric field $E_x$ of 0.05--0.4~V/nm in the $x$ direction
(linear response to the applied force was verified
for all reported results). Starting from an initial
random configuration with zero total linear momentum, simulations
were carried out for roughly 10~ns, with statistics collected
only after the steady state had been reached (typically 1~ns).
Surface charge densities $\Sigma$ of 0, $\pm 0.031$,
and $\pm 0.062$~C/m$^2$ and electrolyte concentrations of
approximately 0.2 and 1~M (8 and 40 ion pairs, respectively,
for $\Sigma=0$) were studied.

\begin{figure}
\includegraphics*[width=8.5cm]{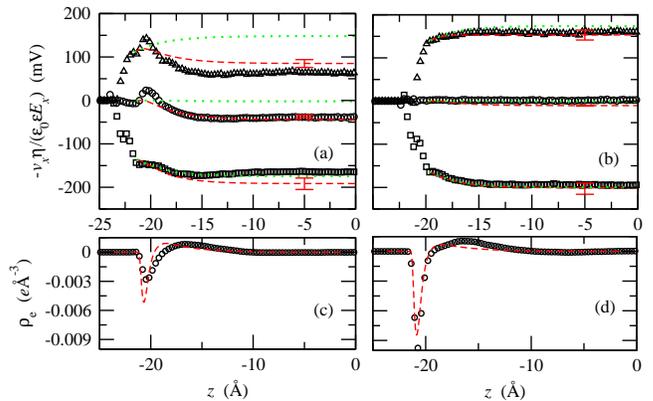}
\caption{\label{fig:vel_den}{\it Top}: Velocity profiles in a hydrophobic chanel for $\Sigma =
-0.062$, 0, and  $+0.062~\text{C/m}^2$ (from bottom to top) with
(a) $[\text{NaI}]\approx 1$~M and (b) $[\text{NaCl}]\approx 1$~M.
The simulation results (symbols) are compared with the 
resolution of the modified PB equation using (see text for details)
the Step-Polarization model (dashed),
and the Step-Polarization model with 
$U_{\text{hyd}}^{\pm}=0$ (dotted). Typical error bars for the theoretical curves
are shown. Error bars in the simulated velocities are roughly the
size of the points. 
{\it Bottom}:  ionic charge density profile $\rho_{\text{e}}(z)$ for
$[\text{NaI}]\approx 1$~M with (c) $\Sigma = 0$ and (d) $\Sigma =
+0.062~\text{C/m}^2$. The symbols are from simulation. For (c) and
(d) the lines are solutions of the modified PB equation with the Step-Polarization model.}
\vspace{-0.5cm}
\end{figure}

\begin{figure}
\includegraphics*[width=8cm,height=!]{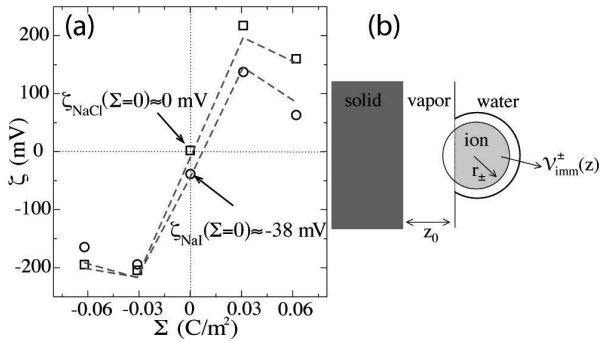}
\vspace{-0.5cm}
\caption{\label{fig:zeta} (a) Zeta potential of the hydrophobic surface versus
 surface charge for 1-M solutions of NaI and NaCl 
 (simulation: NaI -- circles; NaCl -- squares).
The lines are solutions of the Stokes equation with
$\rho_{\text{e}}$ from solving PB equation using the Step-Polarization model (dashed) (see text).
Error bars in the simulated $\zeta$ are roughly the
size of the points. (b) Schematic of an ion at solid--liquid
interface, illustrating the origin of $U_{\text{hyd}}^{\pm}$ and its calculation.
}
\vspace{-0.5cm}
\end{figure}

The measured velocity $v_x(z)$ is shown in Fig.~\ref{fig:vel_den}
for the 1-M solutions in a hydrophobic chanel with $\Sigma=$ 0 and $\pm 0.062$~C/m$^2$; it
has been scaled by the bulk viscosity $\eta$, bulk dielectric
constant $\epsilon$, and applied electric field $E_x$ for ease of
comparison with the zeta potential, defined in terms of the
velocity in the channel center as $\zeta \equiv -\eta v_x(0)/
\left(\epsilon_0\epsilon E_x\right)$, where $\epsilon_0$ is the
vacuum permittivity. For $\epsilon$, we used the dielectric
constant of pure SPC/E water under similar thermodynamic
conditions, $\epsilon_{\text{w}} = 68$~\cite{hochtl04}. The zeta
potentials for all the surface charges  are
given in Fig.~\ref{fig:zeta}.

Figures~\ref{fig:vel_den} and ~\ref{fig:zeta} clearly show the
sensitivity of the EO flow to anion type, particularly
for the neutral and positively charged surfaces. (The flow for the
negatively charged surfaces is dominated by the excess of cations,
Na$^+$ in all simulations). A noteworthy point is the measurement of a non-zero $\zeta$ potential
($\zeta\simeq -40$ mV) for the
neutral hydrophobic channel with a solution of NaI, even though the total electrostatic force
exerted on the charge-neutral fluid is zero. This is in strong contrast to the
traditional theory of the electric double layer~\cite{hunter01}, though
observed in previous experiments~\cite{dukhin05,petrache06}
and computer simulations~\cite{joseph06}.
By contrast, $\zeta$ for NaCl in the same channel was negligible.
Our measured $\zeta$ potentials of 0 and -38~mV respectively
for 1-M NaCl and NaI are consistent with experimental surface
potentials of roughly 0 and -20~mV respectively
for vapor--liquid interfaces of the same solutions~\cite{jarvis68};
$\zeta$ for NaI is also of similar magnitude to
the value of -9~mV measured by electrophoresis of neutral
liposomes in 1-M KI~\cite{petrache06},
for which ion-specific effects should be smaller due to the
greater similarity in size of K$^+$ and I$^-$ compared with Na$^+$ and I$^-$.
Although not shown in Fig.~\ref{fig:zeta}, we
find that the zeta potential for $\Sigma=0$ is not sensitive to electrolyte concentration. Also,
$\zeta$ was insignificant for NaI in the neutral
hydrophilic channel. 

The anomalous result for the uncharged walls can be understood in
terms of continuum hydrodynamics, in which the EO flow is
described by the Stokes equation~\cite{hunter01},
$\frac{\text{d}^2v_x(z)}{\text{d}z^2} =
-\frac{E_x}{\eta} \rho_{\text{e}}(z)$,
where
$\rho_{\text{e}}(z) = e\left[\rho_+(z) - \rho_-(z)\right]$
is the total charge density due to cations and anions of
number density $\rho_{\pm} \left(z \right)$ and $e$ is the elementary charge.
Exploiting the symmetry of our system about $z=0$ and
integrating the Stokes equation twice with boundary conditions (BCs)
$v_x\vert_{z=z_{\text{h}}} =
 b\frac{\text{d}v_x}{\text{d}z}\vert_{z=z_{\text{h}}}$ and
$\frac{\text{d}v_x}{\text{d}z}\vert_{z=0} = 0$, where $b$ is the slip
length applied at the hydrodynamic boundary $z_{\text{h}}$ \cite{Joly}, gives
\begin{equation}
\zeta \equiv -\frac{\eta v_x(0)}
{\epsilon_0\epsilon_{\text{w}} E_x} =
-\frac{1}{\epsilon_0\epsilon_{\text{w}}}
\int_{z_{\text{h}}}^0 \text{d}z' (z'-z_{\text{h}}+b)\rho_{\text{e}}\left(z'\right).
\label{eqn:zeta}
\end{equation}
According to Eq.~(\ref{eqn:zeta}), $\zeta$ is proportional to the
first moment of the charge distribution $\rho_{\text{e}}$ relative to an
origin at the shear plane, $z_{\text{s}} = z_{\text{h}} - b$
(the plane where the non-slip BC applies).
Unless $\rho_{\text{e}}(z) = 0$ everywhere, this quantity will
generally be non-zero even if the total charge,
$\int_{z_{\text{h}}}^0 \text{d}z'\rho_{\text{e}}(z') $, is zero,
as is the case for NaI near the uncharged
hydrophobic wall due to the differing propensities of
Na$^+$ and I$^-$ for the surface.
It has been suggested that such a non-zero $\zeta$ potential occurs
for some non-charged surfaces due to ion-specific ``binding''~\cite{petrache06},
to the presence of an immobile interfacial layer of charge~\cite{dukhin05}, or more
generally to a reduced mobility in the interfacial layer~\cite{joseph06}.
In contrast, our present results show that $\zeta$ will be non-zero even if all of
the charge is fully mobile. Another interesting consequence of Eq.~(\ref{eqn:zeta})
is that, as long as $b$ is finite,
surface slippage makes no contribution to the velocity
of a charge-neutral fluid containing only mobile charge:  i.e. the system
behaves as if $b=0$ and the flow is independent of the solid--fluid friction.
As a matter of fact, the global fluid neutrality
requires the wall-to-fluid force to vanish in the steady state which imposes
both the slip velocity and the velocity gradient to vanish at the wall
(see fig.~\ref{fig:vel_den} for $\Sigma=0$). In this respect, the case where $b$ is infinite
appears singular as the velocity at $z_{\text{h}}$ need not vanish: the velocity is determined
by momentum conservation, as momentum cannot be transferred to the frictionless surface, and
in the end no net flow is achieved (not shown).

So far, we have presented a general explanation for the observed
ion-specific electrokinetic effects; however, of additional practical
value would be a model capable of quantitatively predicting
the ion density and EO velocity profiles. With this aim, we have sought
to construct the minimal physically accurate model
for $\rho_{\text{e}}$ for use in the Stokes equation
for $v_x$.
To obtain $\rho_{\text{e}}$, we solved the one-dimensional
Poisson equation~\cite{hunter01},
$\frac{\text{d}}{\text{d}z} \left[-\epsilon_0
\frac{\text{d}}{\text{d}z}  V(z)
+ P(z) \right] = \rho_{\text{e}}(z)$,
with Neumann BCs applied at the position
$z_{\text{w}}$ of the surface charge
and a mean-field approximation for the ion densities,
$\rho_{\pm}(z)= \rho_0 \exp\left\{-\beta\left[\pm eV(z)
+ U_{\text{ext}}^{\pm}(z)  \right]\right\}$,
where $\rho_0$ is the bulk ion density and $U_{\text{ext}}^{\pm}$
is an external potential acting on the ions due to interactions
other than the electrical potential $V$. For the polarization of
the medium, $P$, we 
assumed $\epsilon
(z)$ to display a step-function behavior at the vapor--liquid interface
(from $\epsilon_0$ to $\epsilon_{\text{w}}$) so that
$P(z) = -\epsilon_0\left[\epsilon_{\text{w}} - 1 \right]
\frac{\text{d}V(z)}{\text{d}z}$, for $z \geq z_0$ and $P=0$ otherwise, with
$z_0$ the position of the first peak in the simulated water oxygen
density distribution function (``Step-Polarization'' (SP) model).

For the external potential, we used the sum of three components:
$U_{\text{ext}}^{\pm} = U_{\text{image}}^{\pm}
+ U_{\text{wall}}^{\pm} + U_{\text{hyd}}^{\pm}$.
The first two terms, $U_{\text{image}}^{\pm}$ and
$U_{\text{wall}}^{\pm}$, are respectively the image potential
acting on the ions due to the dielectric interface at $z_0$
(Eq.~(3) in~\cite{bostrom05}) and the ion--solid LJ interaction,
obtained by integrating the inter-particle LJ interaction over a
uniform density $\rho_{\text{s}}$ of solid atoms occupying the
$z \leq z_{\text{w}}$ half-plane. The final term,
$U_{\text{hyd}}^{\pm}$, is the free energy to create an ion-sized cavity
in the fluid, i.e. to solvate a solute with no attraction to
the solvent. This hydrophobic solvation energy has generally been ignored
in calculations
of interfacial ion densities, since it is negligible compared with
electrostatic interactions for typical small ions
like Na$^+$ or Cl$^-$.
We took $U_{\text{hyd}}^{\pm}$ to be proportional to the volume $\mathcal{V}_{\text{imm}}^{\pm}$ of the
ion immersed in the liquid in the $z \geq z_0$ half-plane
(see Fig.~\ref{fig:zeta}b):
\begin{equation}
 U_{\text{hyd}}^{\pm} (z)
= C_0\left(\mathcal{V}_{\text{imm}}^{\pm}(z)-\mathcal{V}_{\text{ion}}^{\pm}\right),
\label{eqn:Uhyd}
\end{equation}
with $\mathcal{V}_{\text{ion}}^{\pm}$ the total volume of the ion of solvent-excluded radius $r_{\pm}$.
We took  $r_{\pm}$ from
bulk simulations of ions in water as the radius at which the
ion--water radial distribution function fell to $1/e$ of its bulk
value (2.24, 2.98, and 3.73~\AA\ respectively for Na$^+$, Cl$^-$,
and I$^-$). For the proportionality constant, $C_0 = 2.8 \times
10^8~\text{J/m}^3$, we used the solvation free energy per unit
volume measured under similar thermodynamic conditions in
simulations of hard-sphere solutes of radius 0--5 \AA\ in SPC/E
water~\cite{huang01}. We omitted a final plausible term in
$U_{\text{ext}}^{\pm}$, the Born solvation energy
$U_{\text{Born}}^{\pm}$ for charging the ion-sized cavity, as we
found it made little difference to our results, at least using a
relatively simple expression employed by Bostr\"om et
al.~\cite{bostrom05}.

The Stokes equation was solved using the calculated
$\rho_{\text{e}}(z)$ and $\eta$ and $b$ measured independently in
Poiseuille and Couette flow simulations
respectively~\cite{pois_cou}. For $\Sigma = 0$, we used $b = 0$,
as justified above. BCs were applied in all cases at the position
$z_{\text{h}}$ of the first peak in the simulated water oxygen density
distribution function~\cite{z_h}. As a test of the validity of the
continuum hydrodynamic description, we solved the Stokes equation
using the exact $\rho_{\text{e}}(z)$ from our simulations and
found almost perfect agreement with the simulated velocity
profiles (not shown). Both the charge density profile $\rho_{\text{e}}(z)$ and
velocity profiles calculated from the modified
PB theory described above with the full $U_{\text{ext}}^{\pm}$ are 
in good agreement with the simulated results, as shown in Fig.~\ref{fig:vel_den}.
The resulting prediction for the $\zeta$ potential reproduces very well,
both qualitatively and quantitatively, the simulation results, as shown in
Fig.~\ref{fig:zeta}.
It should be noted that the non-monotonic behavior of $\zeta$ as a
function of $\Sigma$ in Fig.~\ref{fig:zeta}
is due to the decrease in the slip length $b$ with $\Sigma$.
Note that it is possible to replace the SP model for $P(z)$ by the exact value
of the polarization (the gradient of which is equal to minus the charge density
due to water in our simulations): doing so yields an even better agreement of the predicted
$\zeta$ potentials with simulation results (not shown) but at the expense of using the
simulated water charged density profile as an input.
Finally, when
$U_{\text{hyd}}^{\pm}$ is neglected the calculated ion
density profiles and velocities are significantly wrong for the
neutral and positively charged surfaces (see Fig. \ref{fig:vel_den}), 
pointing
to the crucial role of the hydrophobic solvation energy. This is not an
issue for the negatively charged surfaces, since the flow is
dominated by Na$^+$, for which $U_{\text{hyd}}^{+}$ is negligible.
Although not shown, we found that the conventional theory of
the electric double layer, which assumes $\epsilon(z) =
\epsilon_{\text{w}}$ and $U_{\text{ext}}^{\pm}=0$ everywhere,
performed very poorly in almost all cases.

In summary, we have shown that anomalous electrokinetic effects
such as non-zero $\zeta$ potentials for uncharged surfaces are generic
features of EO flow in hydrophobic channels when the
dissolved cation and anion differ substantially in size.
We have also developed a simple model, comprising continuum hydrodynamic
equations and a modified PB description for the ion densities, which
accurately predicts the simulated flow profiles. We have found that
the incorporation in the model of an ion-size-dependent hydrophobic
solvation energy, which favors interfacial enhancement of large ions,
is crucial to reproducing the ion-specific effects observed in the
simulations. Such an analytic theory, which is able to capture the subtle
and complex effects of the interfacial specificity of ions, provides a very useful framework
for the modeling of biological systems, for which Hofmeister series are ubiquitous \cite{Ninham}.

This work is supported by ANR PNANO, Nanodrive.


\begin{thebibliography}{99}
\expandafter\ifx\csname natexlab\endcsname\relax\def\natexlab#1{#1}\fi
\expandafter\ifx\csname bibnamefont\endcsname\relax
  \def\bibnamefont#1{#1}\fi
\expandafter\ifx\csname bibfnamefont\endcsname\relax
  \def\bibfnamefont#1{#1}\fi
\expandafter\ifx\csname citenamefont\endcsname\relax
  \def\citenamefont#1{#1}\fi
\expandafter\ifx\csname url\endcsname\relax
  \def\url#1{\texttt{#1}}\fi
\expandafter\ifx\csname urlprefix\endcsname\relax\def\urlprefix{URL }\fi
\providecommand{\bibinfo}[2]{#2}
\providecommand{\eprint}[2][]{\url{#2}}

\bibitem{Chandler} D. Chandler, Nature {\bf 437}, 640 (2005)
\bibitem{Lauga} E. Lauga, M. Brenner, H. Stone, Handbook of Experimental Fluid Dynamics (Springer, 2006)
\bibitem{Joly}ÊL. Joly, C. Ybert, E. Trizac, L. Bocquet, Phys. Rev. Lett. {\bf 93}, 257805 (2004).

\bibitem{Attard} P. Attard, Adv. Coll. Int. Sci. {\bf 104}, 75 (2003)
\bibitem[{\citenamefont{Netz}(2004)}]{netz04}
\bibinfo{author}{\bibfnamefont{R.~R.} \bibnamefont{Netz}},
  \bibinfo{journal}{Curr. Opin. Coll. Int. Sci.}
  \textbf{\bibinfo{volume}{9}}, \bibinfo{pages}{192} (\bibinfo{year}{2004}).

\bibitem[{\citenamefont{Ghosal et~al.}(2005)\citenamefont{Ghosal, Hemminger,
  Bluhm, Mun, Hebenstreit, Ketteler, Ogletree, Requejo, and
  Salmeron}}]{ghosal05}
\bibinfo{author}{\bibfnamefont{S.}~\bibnamefont{Ghosal}},
  \bibinfo{author}{{\it et al.}},
  \bibinfo{journal}{Science} \textbf{\bibinfo{volume}{307}},
  \bibinfo{pages}{563} (\bibinfo{year}{2005}).

\bibitem[{\citenamefont{Vrbka et~al.}(2004)\citenamefont{Vrbka, Mucha, Minofar,
  Jungwirth, Brown, and Tobias}}]{vrbka04}
\bibinfo{author}{\bibfnamefont{L.}~\bibnamefont{Vrbka}},
  \bibinfo{author}{{\it et al.}},
  \bibinfo{journal}{Curr. Opin. Coll. Int. Sci.}
  \textbf{\bibinfo{volume}{9}}, \bibinfo{pages}{67} (\bibinfo{year}{2004}).

\bibitem[{\citenamefont{Hunter}(2001)}]{hunter01}
\bibinfo{author}{\bibfnamefont{R.~J.} \bibnamefont{Hunter}},
  \emph{\bibinfo{title}{Foundations of Colloid Science}}
  (\bibinfo{publisher}{Oxford University Press, Oxford}, \bibinfo{year}{2001}),
  \bibinfo{edition}{2nd} ed.

\bibitem[{\citenamefont{Bostr{\"o}m et~al.}(2005)\citenamefont{Bostr{\"o}m,
  Kunz, and Ninham}}]{bostrom05}
\bibinfo{author}{\bibfnamefont{M.}~\bibnamefont{Bostr{\"o}m}},
  \bibinfo{author}{\bibfnamefont{W.}~\bibnamefont{Kunz}}, \bibnamefont{and}
  \bibinfo{author}{\bibfnamefont{B.~W.} \bibnamefont{Ninham}},
  \bibinfo{journal}{Langmuir} \textbf{\bibinfo{volume}{21}},
  \bibinfo{pages}{2619} (\bibinfo{year}{2005}).

\bibitem[{\citenamefont{Squires and Quake}(2005)}]{squires05}
\bibinfo{author}{\bibfnamefont{T.} \bibnamefont{Squires}}, 
  \bibinfo{author}{\bibfnamefont{S.} \bibnamefont{Quake}},
  \bibinfo{journal}{Rev. Mod. Phys.} \textbf{\bibinfo{volume}{77}},
  \bibinfo{pages}{977} (\bibinfo{year}{2005}).

\bibitem{Ninham} M. Bostr\"om, D.R.M. Williams, B. W. Ninham, Phys. Rev. Lett. {\bf 87}
168103 (2001).

\bibitem[{\citenamefont{Plimpton}()}]{lammps}
\bibinfo{author}{\bibfnamefont{S.~J.} \bibnamefont{Plimpton}},
  \bibinfo{note}{{J}. Comput. Phys. \textbf{117}, 1 (1995); \texttt{LAMMPS}:
  \texttt{http://lammps.sandia.gov}.}

\bibitem[{\citenamefont{Koneshan et~al.}(1998)\citenamefont{Koneshan, Rasaiah,
  Lynden-Bell, and Lee}}]{koneshan98}
\bibinfo{author}{\bibfnamefont{S.}~\bibnamefont{Koneshan}},
  \bibinfo{author}{\bibfnamefont{J.~C.} \bibnamefont{Rasaiah}},
  \bibinfo{author}{\bibfnamefont{R.~M.} \bibnamefont{Lynden-Bell}},
  \bibnamefont{and} \bibinfo{author}{\bibfnamefont{S.~H.} \bibnamefont{Lee}},
  \bibinfo{journal}{J. Phys. Chem. B} \textbf{\bibinfo{volume}{102}},
  \bibinfo{pages}{4193} (\bibinfo{year}{1998}).

\bibitem[{\citenamefont{Archontis and Leontidis}(2006)}]{archontis06}
\bibinfo{author}{\bibfnamefont{G.}~\bibnamefont{Archontis}} \bibnamefont{and}
  \bibinfo{author}{\bibfnamefont{E.}~\bibnamefont{Leontidis}},
  \bibinfo{journal}{Chem. Phys. Lett.} \textbf{\bibinfo{volume}{420}},
  \bibinfo{pages}{199} (\bibinfo{year}{2006}).

\bibitem[{\citenamefont{H{\"o}chtl et~al.}(1998)\citenamefont{H{\"o}chtl,
  Boresch, Bitomsky, and Steinhauser}}]{hochtl04}
\bibinfo{author}{\bibfnamefont{P.}~\bibnamefont{H{\"o}chtl}},
  \bibinfo{author}{\bibfnamefont{S.}~\bibnamefont{Boresch}},
  \bibinfo{author}{\bibfnamefont{W.}~\bibnamefont{Bitomsky}}, \bibnamefont{and}
  \bibinfo{author}{\bibfnamefont{O.}~\bibnamefont{Steinhauser}},
  \bibinfo{journal}{J. Chem. Phys.} \textbf{\bibinfo{volume}{109}},
  \bibinfo{pages}{4927} (\bibinfo{year}{1998}).

\bibitem[{\citenamefont{Dukhin et~al.}(2005)\citenamefont{Dukhin, Dukhin, and
  Goetz}}]{dukhin05}
\bibinfo{author}{\bibfnamefont{A.}~\bibnamefont{Dukhin}},
  \bibinfo{author}{\bibfnamefont{S.}~\bibnamefont{Dukhin}}, \bibnamefont{and}
  \bibinfo{author}{\bibfnamefont{P.}~\bibnamefont{Goetz}},
  \bibinfo{journal}{Langmuir} \textbf{\bibinfo{volume}{21}},
  \bibinfo{pages}{9990} (\bibinfo{year}{2005}).

\bibitem[{\citenamefont{Petrache et~al.}(2006)\citenamefont{Petrache, Zemb,
  Belloni, and Parsegian}}]{petrache06}
\bibinfo{author}{\bibfnamefont{H.~I.} \bibnamefont{Petrache}},
  \bibinfo{author}{\bibfnamefont{T.}~\bibnamefont{Zemb}},
  \bibinfo{author}{\bibfnamefont{L.}~\bibnamefont{Belloni}}, \bibnamefont{and}
  \bibinfo{author}{\bibfnamefont{V.~A.} \bibnamefont{Parsegian}},
  \bibinfo{journal}{Proc. Natl. Acad. Sci. USA} \textbf{\bibinfo{volume}{103}},
  \bibinfo{pages}{7982} (\bibinfo{year}{2006}).

\bibitem[{\citenamefont{Joseph and Aluru}(2006)}]{joseph06}
\bibinfo{author}{\bibfnamefont{S.}~\bibnamefont{Joseph}} \bibnamefont{and}
  \bibinfo{author}{\bibfnamefont{N.~R.} \bibnamefont{Aluru}},
  \bibinfo{journal}{Langmuir} \textbf{\bibinfo{volume}{22}},
  \bibinfo{pages}{9041} (\bibinfo{year}{2006}).

\bibitem[{\citenamefont{Jarvis and Scheiman}(1968)}]{jarvis68}
\bibinfo{author}{\bibfnamefont{H.~L.} \bibnamefont{Jarvis}} \bibnamefont{and}
  \bibinfo{author}{\bibfnamefont{M.~A.} \bibnamefont{Scheiman}},
  \bibinfo{journal}{J. Phys. Chem.} \textbf{\bibinfo{volume}{72}},
  \bibinfo{pages}{74} (\bibinfo{year}{1968}).
\bibitem[{\citenamefont{Huang and Chandler}(2001)}]{huang01}
\bibinfo{author}{\bibfnamefont{D.~M.} \bibnamefont{Huang}},
\bibinfo{author}{\bibfnamefont{P.L.}~\bibnamefont{Geissler}} \bibnamefont{and}
  \bibinfo{author}{\bibfnamefont{D.}~\bibnamefont{Chandler}},
  \bibinfo{journal}{J. Phys. Chem. B} \textbf{\bibinfo{volume}{105}},
  \bibinfo{pages}{6704} (\bibinfo{year}{2001}).
  
\bibitem[{poi()}]{pois_cou}
\bibinfo{note}{We obtained $\eta$ from independent Poiseuille flow simulations : 
$\eta_{\text{w}}= 0.65\pm 0.06$, $0.75\pm 0.04$, and $0.89\pm
  0.04$~mPa~s\ respectively for pure SPC/E water, 1-M NaI, and 1-M NaCl. 
  From Couette flow
  simulations, 
  we obtained $b$ as the distance beyond $z_{\text{h}}$
  at which the linearly extrapolated fluid velocity was equal to the wall
  velocity: $b\approx$ 30~\AA\ and 10~\AA\ (errors roughly $\pm 1.1$ and $\pm
  0.4$~\AA) respectively for $\Sigma = \pm 0.031$ and $\pm 0.062~\text{C/m}^2$,
  with slight dependence on electrolyte type for 1-M solutions.}

\bibitem[{z_h()}]{z_h}
\bibinfo{note}{Changing the definition of $z_{\text{h}}$ results in no change
  to $v_x \left(z\right)$ if $\rho_{\text{e}}\left(z \right) = 0$ for $ z <
  z_{\text{h}}$; we also found our chosen definition resulted in consistent
  values for $b$ from Couette and Poiseuille flow simulations.}

\end{thebibliography}

\end{document}